\documentstyle[12pt]{article}
\begin{document}

\baselineskip=0.5truecm
\title{
Recent Results on the Decay of Metastable Phases
}
\author{
\bf{Per~Arne~Rikvold and Bryan~M.~Gorman}\\
\\
Supercomputer Computations Research Institute, \\
Department of Physics, and\\
Center for Materials Research and Technology, \\
Florida State University, Tallahassee, FL 32306-4052, U.S.A.\\
}
\maketitle
\begin{abstract}
We review some aspects of current knowledge regarding the decay of metastable
phases in many-particle systems. In particular we emphasize recent theoretical
and computational developments and numerical results
regarding homogeneous nucleation and growth in kinetic Ising
and lattice-gas models. An introductory discussion of the droplet
theory of homogeneous nucleation is followed by a discussion of
Monte Carlo and transfer-matrix methods commonly used for numerical study
of metastable decay, including some new algorithms.
Next we discuss specific classes of systems.
These include a brief discussion of recent progress for fluids, and more
exhaustive considerations of ferromagnetic Ising models ({\it i.e.},
attractive lattice-gas models) with weak long-range interactions and with
short-range interactions. Whereas weak-long-range-force (WLRF)
models have infinitely long-lived metastable phases in the
infinite-range limit, metastable phases
in short-range-force (SRF) models eventually decay, albeit extremely slowly.
Recent results on the finite-size scaling of metastable lifetimes in SRF
models are reviewed, and it is pointed out that such effects may be
experimentally observable.
\end{abstract}

\clearpage


\section{Introduction}
\label{sec1}

Metastable phases are common in nature, and some of them have such extremely
long lifetimes that they are practically indistinguishable from equilibrium
phases. For example, even though the stable phase of carbon at room
temperature and atmospheric pressure
is graphite, the claim that diamonds are forever is
not likely to be challenged on physical grounds.
Other examples of metastable phases
with average lifetimes that are measured in milliseconds to
days, rather than in millions or billions of years, are
supercooled or supersaturated fluids
[1--32],
ferroelectrics
[33--38],
the small single-domain ferromagnetic particles important in
paleomagnetism and high-density recording media
[39--49],
and vortex states in superconductors \cite{OVCH93}. Further
possible examples are the supercooled quark/gluon plasma associated with the
QCD confinement transition \cite{KAJA92,HACK92}
and the ``false vacuum'' associated
with the electroweak transition
[53--57],
both of which may have played
important roles in the early development of the universe.

To clarify what we understand by a metastable phase, it is hard to
improve on the empirical descriptions due to
Penrose and Lebowitz \cite{PENR71,PENR79} and Sewell \cite{SEWE80}. \\
({\it i}) The free energy of the system is not fully minimized.\\
({\it ii}) Only one thermodynamic phase is present, and for sufficiently
small and slow perturbations the usual laws of reversible thermodynamics
apply. \\
({\it iii}) A system starting in the metastable phase
is likely to take a long time to escape. \\
({\it iv}) Escape is irreversible: once out of the metastable phase,
the system is extremely unlikely to return.

We concentrate on systems for which the order parameter
is a nonconserved scalar (Model A in the Hohenberg-Halperin
scheme of dynamic universality classes \cite{HOHE77})
so that the metastability is
imposed by an applied external field. (The closely related phenomenon of
hysteresis imposed by an oscillating
field is discussed by Acharyya and Chakrabarti elsewhere in this
volume \cite{ACHA94}.) Even so, many of the phenomena we
discuss can be generalized to systems with a multidimensional
and/or conserved order parameter .

The wide variety of contexts in which metastability has been studied, often
independently, makes it difficult to present a discussion that is equally
accessible to all potentially interested readers. In this review we have
chosen to use the language of Ising models.
The most important reason for this choice is our
belief that the present theoretical
understanding of metastable phases and their modes of decay is most
highly developed in applications to the relatively simple
kinetic Ising models and
their equivalent formulations as lattice-gas models.
A secondary reason is the high symmetry of the Ising formulation.
Throughout this paper we therefore use Ising ferromagnets as
generic examples. Below we define these models and give the standard
transformations that will enable the reader to convert our results to
the lattice-gas language appropriate to,
{\it e.g.}, fluids, binary mixtures, and adsorbate systems.

An Ising ferromagnet is defined by a
Hamiltonian ({\it i.e.}, an energy functional),
\begin{equation}
\label{eqISING}
{\cal H} = -  \frac{1}{2} \sum_{i,j} J_{i,j} s_i s_j - H \sum_i s_i \;,
\end{equation}
where $s_i \! = \! \pm 1$ (``up'' or ``down'') is a binary variable,
or ``spin,'' at site $i$, $H$ is the applied field, and the sums run over
all $\cal N$ sites on a $d$-dimensional lattice.
The interaction energies $J_{i,j}$ are positive
and symmetric under interchange of the site indices $i$ and $j$, and
without loss of generality we set $J_{i,i}$=0.
By requiring that $\sum_j J_{i,j} \! \equiv \! zJ$ is independent of $i$
we ensure the spatial homogeneity of the energy functional.
We also note that $\cal H$ is invariant under the transformation
$\{ s_i \rightarrow -$$s_i, \; H \rightarrow -$$H \}$.
The order parameter conjugate to $H$
is the ``magnetization'' or ``polarization'',
\begin{equation}
\label{ISINGm}
m = {\cal N}^{-1} \sum_{i} s_i \;.
\end{equation}
As is well known \cite{STAN71}, this model is equivalent to a two-state
lattice-gas model with local concentration variables $c_i$=0 or~1
(``empty'' or ``occupied''), for which Eq.~(\ref{eqISING}) takes the form
\begin{equation}
\label{eqLG}
{\cal H} = -  \frac{1}{2} \sum_{i,j} \Phi_{i,j} c_i c_j - \mu \sum_i c_i
+ \frac{{\cal N}}{2} \left( \mu - \frac{1}{2} \mu_0 \right)
\equiv {\cal H}_{\rm LG}
+ \frac{{\cal N}}{2} \left( \mu - \frac{1}{2} \mu_0 \right)
\;,
\end{equation}
where the quantities appearing in the two equivalent formulations of the
Hamiltonian, Eq.~(\ref{eqISING}) and Eq.~(\ref{eqLG}), are linked
by the transformations
\begin{eqnarray}
\label{eqISLG}
c_i        &=& (s_i \! + \! 1)/2 \nonumber\\
\Phi_{i,j} &=& 4J_{i,j}          \nonumber\\
\mu        &=& 2H \! + \! \mu_0   \;.
\end{eqnarray}
Here $\Phi_{i,j}$ are attractive lattice-gas interaction energies and
$\mu$ is the chemical potential, whose value at coexistence ({\it i.e.},
for $H$=0) is $\mu_0$=$-$$2zJ$.
The chemical potential is related to the (osmotic) pressure $p$
as $\mu \! - \! \mu_0 = k_{\rm B}T \ln (p/p_0)$,
where $k_{\rm B}T$ is Boltzmann's constant times the absolute temperature,
and $p_0$ is the pressure at coexistence.
The order parameter conjugate to $\mu$ is the density,
\begin{equation}
\label{LGm}
\rho = {\cal N}^{-1} \sum_{i} c_i = (m+1)/2 \;.
\end{equation}

The energy functional $\cal H$ is not a quantum-mechanical Hamiltonian and
does not impose a unique dynamic on the system. When one uses an Ising or
lattice-gas model to describe the kinetics of a physical system, one must
therefore define a specific stochastic dynamic to simulate the interactions
with the system's surroundings. Usually it is physically most
realistic to choose a dynamic which is {\it local\/} in the sense that it
only allows transitions involving a single site or a pair of nearest-neighbor
sites, and in this review we mostly limit our attention to such dynamics.
Simple examples are the Metropolis \cite{METR53} and Glauber \cite{GLAU63}
dynamics, which are among the algorithms we consider in the context of Monte
Carlo simulations in Sec.~\ref{Sec-MC}. (Under somewhat restrictive
conditions, the Glauber dynamic has been obtained
from the quantum mechanical equations of
motion for a system of distinguishable spin-1/2 particles weakly coupled
to an infinitely large quantum reservoir \cite{MART77}.)

The Ising (lattice-gas) model below its critical temperature
and in the absence of an applied field (with $\mu$=$\mu_0$)
has two coexisting, ordered phases of equal free energy: one with
positive magnetization (high density) and one with negative magnetization
(low density). This degeneracy is lifted by applying
a nonzero field (changing $\mu$ away from $\mu_0$):
the phase whose magnetization is parallel to the field
(for which $\rho(\mu)$$-$$\rho(\mu_0)$ has the same sign as
$\mu$$-$$\mu_0$) becomes the unique equilibrium phase, and the one with
the opposite magnetization (``wrong'' density) becomes metastable.

The system can be prepared in the metastable phase by
equilibrating it in a nonzero field that is then instantaneously reversed.
Although the system is no longer in equilibrium immediately after the field
reversal, it is nevertheless stable against small fluctuations, and its
thermodynamic properties are similar to those it would have in the equilibrium
phase. This is because configurations obtained by flipping small clusters of
neighboring spins cost more free energy
by introducing an interface between previously parallel spins
than they gain by lowering the field energy.
In order for this metastable phase to decay, it is therefore necessary that a
cluster is created that is sufficiently large
for the free energy gained by aligning more spins
with the field to just outweigh the cost of breaking the necessary
extra bonds. Such a fluctuation corresponds to a local maximum in the
free-energy landscape and is usually called a
``critical nucleus'' or a ``critical droplet''.
Once randomly created through a thermal fluctuation in the metastable phase,
it is likely to grow further. During this growth period
the now ``supercritical'' droplet incorporates spins
from the metastable phase into a growing, and soon macroscopic, domain of the
equilibrium phase. The timescale for the creation of a critical droplet is in
general much longer than that characteristic of the subsequent growth.

Real systems are of course more complicated than the simple picture
discussed above.
Nevertheless, it covers the essential physics of nucleation in a variety
of situations.
Some of the additional complications that arise in the study of metastability
in fluids with continuous degrees of freedom
are briefly discussed in Sec.~\ref{Sec-LV}.

The nucleation mechanism described above, which
is usually known as {\it homogeneous} or {\it thermally activated} nucleation,
will be the focus of our attention.
It is the dominant mechanism in systems that are free of defects, or in which
the defect concentration is low, or
in which the defects are much smaller than the
critical droplet size. Whenever this is not the case, heterogeneous nucleation
on defects or interfaces may be the dominant process for producing
droplets of the equilibrium phase \cite{SHNE94}.
However, the subsequent growth process
depends little on the nucleation mechanism \cite{ISHI71,EVAN45}.

The organization of the remainder of this review is as follows.
Section \ref{Sec-FT} contains a brief review of classical nucleation
theory and some of the general results obtained in the ``post-classical''
period after the 1940's. In Sec.~\ref{Sec-Cmet} we review numerical
Monte Carlo (in Sec.~\ref{Sec-MC}) and transfer-matrix (in Sec.~\ref{Sec-CTM})
methods to study nucleation and metastable decay, primarily in kinetic Ising
models. Following these general sections we consider specific classes of
systems: fluids in Sec.~\ref{Sec-LV}, Ising models with weak long-range
forces (WLRF models)
and their relation to mean-field approximations in Sec.~\ref{Sec-LRF},
and Ising models with short-range forces (SRF models) in Sec.~\ref{Sec-SRF}.
The subsections in Sec.~\ref{Sec-LRF} are Sec.~\ref{Sec-LRFn}, which
contains specific nucleation rates for the ramified critical droplets
characteristic of WLRF models, and Sec.~\ref{Sec-LRFtm}, which
contains numerical results.
The subsections in Sec.~\ref{Sec-SRF} are Sec.~\ref{Sec-SRFn}, which
contains specific nucleation rates for the compact critical droplets
characteristic of SRF models, Sec.~\ref{Sec-KJMA}, in which we
discuss the interplay  between nucleation and the {\it growth} of
supercritical droplets, Sec.~\ref{Sec-FSE}, in which we consider finite-size
effects, and Sec.~\ref{Sec-DL}, in which we consider effects of the discrete
lattice at low temperatures.
Finally, in Sec.~\ref{Sec-D}, we give a concluding summary and discussion.

\section{Droplet Theory of Homogeneous Nucleation}
\label{Sec-FT}

The first scientific description of metastability may well be
Fahrenheit's experiments with supercooled water,
published in 1724 \cite{F1724}, and further observations were reported
during the eighteenth and early nineteenth centuries by several
workers \cite{DUNN69}.
The basis for a theoretical understanding
was laid 150 years after Fahrenheit's paper with van~der Waals' \cite{VDW1873}
and Maxwell's \cite{MAXW1874} early mean-field approaches
and Gibbs' realization that the reversible
work of formation of a droplet of the stable
phase in a metastable background is a ``measure of the stability'' of the
metastable phase
[4--6].
Although Gibbs' work clearly
states the basic equations of what is today known as ``Classical Nucleation
Theory'' (CNT), further development did not occur until the
period from the 1920's through the 1940's.
Gibbs' essentially thermodynamic ({\it i.e.}, {\it static})
approach was then complemented by kinetic considerations by
Volmer and Weber \cite{VOWE26}, Farkas \cite{FARK27},
Becker and D{\"o}ring \cite{BEDO35}, and Zeldovich \cite{ZELD43,FREN46},
whereas Bijl \cite{BIJL38}, Frenkel \cite{FREN39}, and Band \cite{BAND39}
focused on calculating a constrained,
metastable partition function through a
cluster-expansion procedure inspired by that of Mayer \cite{MAYE37}.
The basic result for the nucleation rate
can be written in a Van't~Hoff-Arrhenius \cite{VH1884,AR1889} form,
\begin{equation}
\label{eqCNT}
\Gamma = {\cal P} e^{ - \beta F_{\rm c} } \;,
\end{equation}
where $\cal P$ is a non-exponential prefactor,
$F_{\rm c}$ is the free-energy cost of a critical droplet
(Gibbs' ``measure of stability''), and $\beta$=$1/k_{\rm B}T$ is the
inverse temperature.
(A compact historical sketch is given by Dunning \cite{DUNN69}.
For more recent general reviews of metastability and the kinetics of
first-order phase transitions see, {\it e.g.},
Refs.~[31, 72--74].)

The ``post-classical'' developments in nucleation theory are mainly
efforts to evaluate $F_{\rm c}$ and/or the
prefactor in Eq.~(\ref{eqCNT}) for specific systems.
This observation in itself provides an important insight:
the lifetime of a metastable
state depends crucially on the structure and dynamics
of the excitations through which it decays \cite{FISH90}.

The relevance of the classification scheme for dynamic
critical phenomena due to Hohenberg and Halperin \cite{HOHE77}
to metastable decay is not clear. In addition
to those aspects of the physics that influence both the static and
the dynamic universality,
such as the interaction range and the symmetries and dimensions of
both the system itself and the order parameter, the conservation laws
that influence the dynamic scaling should also be important for the
nucleation rate.

The ``metastable thermodynamics'' that can be observed during the period
{\it before} the decay takes place is almost indistinguishable
from true equilibrium thermodynamics. This has inspired efforts to treat
metastability through suitable generalizations of equilibrium statistical
mechanics.
Notable among these are three papers by Langer
[76--78],
in which he shows that for a wide class of
models, whose dynamics can be described by a Fokker-Planck equation, the
homogeneous nucleation rate (per unit time and volume) may be written as
\begin{equation}
\Gamma = \frac{\beta \kappa}{\pi} | {\rm Im} \widetilde{f} | \;,
\label{eqLANG}
\end{equation}
where ${\rm Im} \widetilde{f}$ is the imaginary part of a
``metastable'' free-energy
density $\widetilde{f}$, which is obtained by analytically continuing
the equilibrium free-energy density $f$ into the metastable phase.
The ``kinetic prefactor'' $\kappa$
contains all dependence on the specific dynamic.
This important work brings together aspects of CNT with
results on droplet theory and analytic continuation from
Andreev \cite{ANDR64} and Fisher \cite{FISH67} and on thermally activated
processes from Landauer and Swanson \cite{LASW61} and Kramers \cite{KRAM40}.

Despite its apparent simplicity and its similarity to the
corresponding relation in quantum mechanics (obtained by replacing
$\widetilde{f}$ by the
energy and $\beta \kappa/ \pi$ by $2/\hbar$) no general proof
of Eq.~(\ref{eqLANG}) exists, and its
domain of validity remains unclear.
Analytic verification for specific models has been reported, {\it e.g.},
by Newman and Schulman \cite{NEWM80} for a Curie-Weiss ferromagnet,
by Roepstorff and Schulman \cite{ROEP84} for an urn model,
by Gaveau and Schulman \cite{GAVE89} for a class of models including the
droplet model of condensation \cite{FREN46,ANDR64,FISH67}, and
by Penrose \cite{PENR94} for the droplet model with Becker-D\"oring dynamics.
A different approach is to assume the validity of Eq.~(\ref{eqLANG})
and calculate $| {\rm Im} \widetilde{f} |$ analytically or numerically.
Again, analytic calculation requires a specific model for the
fluctuations included in the calculation of the analytic continuation
$\widetilde{f}$. Such field-theoretical calculations were done
by Coleman and Callan \cite{COLE77,CALL77} for the ``false vacuum'' in
quantum-field theory,
by B\"uttiker and Landauer \cite{BULA79,BULA81}
for a one-dimensional (1D) overdamped sine-Gordon chain,
by McCraw \cite{MCCR80} for a 1D Kac model with algebraically
decaying interactions,
by G\"unther, Nicole, and Wallace \cite{GNW80}, who generalized
Langer's field-theoretical calculation to arbitrary spatial dimension
(see Sec.~\ref{Sec-SRFn}),
by Zwerger \cite{ZWER85} for a $\phi^4$ field model,
by Cottingham {\it et al.\/} \cite{COTT93} for a model of
bubble formation in fluids,
by Braun \cite{BRAU93,BRAU94} for a model of switching in single-domain
ferromagnetic particles,
and by Klein and Unger \cite{KLEI83,UNGE84}
and Gorman {\it et al.}\ \cite{GORM94,FIIG94},
who used $\phi^3$ field theories to study WLRF systems
near the classical spinodal (see Sec.~\ref{Sec-LRF}).

As may be understood from the heuristic discussion of the decay of a
metastable phase presented in Sec.~\ref{sec1},
the metastable lifetime depends not only on the rate of
nucleation of critical droplets, but equally importantly on the
subsequent rate of growth of the supercritical droplets and on the interplay
between these processes of nucleation and growth. This was
realized by Kolmogorov \cite{KOLM37},
Johnson and Mehl \cite{JOHN39}, and Avrami \cite{AVRAMI}
(KJMA) at about the same time as CNT was being developed,
and a few years later by Evans \cite{EVAN45}.

\section{Computational Methods}
\label{Sec-Cmet}

In contrast to analytical calculations of $\Gamma$ or
$| {\rm Im} \widetilde{f}|$, nonperturbative
numerical methods do not require {\it a priori} knowledge about the critical
excitations. In this section we briefly discuss two classes of such methods:
some varieties of
the well-known Monte Carlo (MC) method for numerical simulation, and
a recently introduced extension of the equilibrium transfer-matrix (TM)
method to also encompass metastable phases.

\subsection{Monte Carlo Methods}
\label{Sec-MC}

Simulations using standard Metropolis or heat-bath dynamics with
spatially local updates
(see, {\it e.g.}, Ref.~\cite{MKB73}) have remained the
computational methods of choice for studying metastable decay in systems with
nonconserved order parameter. The spatial locality of the algorithm
preserves the free-energy barriers that dominate the
nucleation rate, leading to relatively
faithful representations of metastable dynamics in physical systems.
(This is not generally true for cluster algorithms such as the
one due to Swendsen and Wang \cite{SWEN87}, which are used to
accelerate equilibrium simulations near criticality. However, see
further discussion below.)

In equilibrium studies, the choice between the Metropolis \cite{METR53}
and heat-bath algorithms (the latter
also known as the Gibbs sampler or, when applied to the Ising model,
the Glauber \cite{GLAU63} dynamic)
is largely a matter of convenience, and it is often considered
so trivial that it is not even stated explicitly. This habit sometimes has
carried over to dynamical studies where, in our opinion, attention
should be given to finding the dynamic best representing the physical
system to be simulated.
For completeness we give below the probabilities for an allowed
transition from state $x$ to $x'$ for these two algorithms
[101--103].

In the Metropolis algorithm a candidate state is selected according to
a proposal distribution, which vanishes for forbidden transitions.
(The decision about which transitions should be allowed is part of the full
definition of the algorithm. For an Ising system one could for instance
allow single-spin flips,
nonlocal cluster flips as in the Swendsen-Wang \cite{SWEN87} and related
[104--106]
algorithms, or
nearest-neighbor spin exchanges as in the conserved-order-parameter
Kawasaki dynamic \cite{KAWA72}.)
The proposed transition is accepted with probability one if it
leads to a reduction in energy, whereas an increase in energy is accepted
with probability given by a Boltzmann factor, so that the acceptance
probability may be written as
\begin{equation}
\label{eq1aa}
W_{\rm M}( x \rightarrow x' ) =
\min \left\{ 1 ,
\exp \left[ - \! \beta \left( E(x') \! - \! E(x) \right) \right] \right\}
\;.
\end{equation}
In contrast, the heat-bath algorithms accept any state $x'$
to which a transition from $x$ is allowed,
with the equilibrium probability over the set of all
accessible states $x''$:
\begin{equation}
\label{eq1ab}
W_{\rm H}( x \rightarrow x' ) =
\exp \left[ - \! \beta E(x') \right]
\left/ \sum_{\{ x'' \; {\rm accessible \; from \; } x \}}
\exp \left[ - \! \beta E(x'') \right] \right. \;.
\end{equation}

Both the Metropolis and the heat-bath algorithm satisfy detailed balance,
and with ergodicity they eventually converge to thermodynamic equilibrium.
However, the detailed Markov processes generated are not identical.
Beside the choice of the Metropolis or heat-bath transition probability,
more subtle differences between algorithms
may also influence the results of simulations, and even though data obtained
by different local algorithms can often be connected by simply rescaling the
time, the rescaling factor may depend nontrivially on temperature and
distance from the coexistence curve.
For example, in a recent study of metastable decay in the
two-dimensional Ising model with the Metropolis dynamic \cite{RIKV94A},
Rikvold {\it et al.} found that the manner
in which the candidate site for the next spin update was chosen
(sequentially or randomly) affected the observed
field dependence of the kinetic prefactor in Eq.~(\ref{eqLANG}).
Similarly, in a recent study of magnetization relaxation in
the three-dimensional Ising model at the critical point, Ito demonstrated
that the detailed nature
of the finite-size effects depends not only on the choice between
the Metropolis and heat-bath dynamics, but also on whether a sequential or
a checker-board update scheme was used \cite{ITO93}.
In choosing the MC algorithm for a particular study, one should therefore
consider whether the quantities of interest are universal in the sense that
they are the same for all local algorithms, or whether they depend on the
dynamical details of the algorithm used.

For low temperatures and close to coexistence, both the local algorithms
discussed above and the physical dynamic they simulate
spend most of the time creating short-lived, microscopic excitations
in the metastable phase. Because of the smallness of the
subcritical clusters, this is also true for the
modified Swendsen-Wang cluster dynamic used in MC simulations
by Ray, Tamayo, and Wang \cite{RAY90A,RAY90B} and studied theoretically by
Martinelli, Olivieri, and Scoppola \cite{MART91}.
The brute-force way around this problem
is to apply more computer power in the form
of various kinds of supercomputers or special-purpose machines
(as, {\it e.g.}, in Refs.~\cite{RIKV94A,TOMI92A,TOMI92B}),
but inevitably the slowness of the dynamic makes this approach impracticable.
Very recently, two novel MC algorithms have been introduced, which make
simulations deep in the metastable region feasible.
Both methods utilize absorbing Markov chains \cite{IOSI80}.
One of them, introduced by Novotny
[113--115],
generalizes the rejection-free
$n$-fold way algorithm of Bortz, Kalos, and Lebowitz \cite{BORT75} and
achieves CPU-time savings of many orders of magnitude relative to the
local algorithms. This speedup is obtained
without changing the underlying dynamic in any way.
The other method, introduced by Lee {\it et al.} \cite{JLEE94A},
combines absorbing Markov chains with the
Multicanonical method \cite{BERG92}.
The resulting dynamic depends on the microscopic
configurations only through their projection onto the order parameter.
Nevertheless, most qualitative and quantitative features of the dependences
of the metastable lifetimes on field, temperature, and system size
agree with theoretical results and direct simulations for local dynamics.
By comparison with such more traditional methods, this algorithm may help
deepen our understanding of universality in metastable decay.
Preliminary results from both methods are promising
(some are presented in Secs.~\ref{Sec-FSE} and~\ref{Sec-DL}),
and we hope to review further progress in the future.

It is common in dynamical MC simulations of metastable decay
(regardless of the particular
algorithm employed) to study the relaxation of the
order parameter, starting from the metastable phase.
This approach is closely related to the use of
nonequilibrium relaxation functions, introduced by Binder \cite{BIND73A}.
It has been used for SRF models in two
[37, 104, 108, 110, 111, 120--129],
three
[38, 105, 130--132],
and higher \cite{RAY91}
dimensions, and also for WLRF models \cite{PAUL88,HEER84,HEER82,PAUL89}.
Some recent results are presented in Sec.~\ref{Sec-SRF}.

\subsection{The Constrained-Transfer-Matrix Method}
\label{Sec-CTM}

The Constrained-Transfer-Matrix (CTM) method
introduced by one of us \cite{RIKV89}
is particularly suited for numerical
calculation of the analytically continued free-energy
density $\widetilde{f}$ in the field-theoretical
droplet theory discussed in Sec.~\ref{Sec-FT}, without needing a
theoretical model of the critical droplet geometry.
The method extends the usual concept of the transfer matrix (TM) \cite{DOMB60}
to also include constrained nonequilibrium states.
Here we give a brief description of the technique. More detailed discussions
can be found in Refs.~\cite{GORM94,FIIG94,RIKV89,CCAG93,CCAG94A}.

In a standard TM calculation, an $N$$\times$$L$ lattice is
considered in the limit $L$$\rightarrow$$\infty$, and the Hamiltonian
(the energy functional) is written as a sum of layer Hamiltonians,
${\cal H} = \sum_{l=1}^L \bar{\cal H}(x_l,x_{l+1})$,
where $\bar{\cal H}$ depends
only on the configurations $x_l$ and $x_{l+1}$ of two
adjacent layers. The TM is a positive matrix,
\begin{equation}
\label{meth-eq3b}
{\bf T}_0 =
\sum_{x,x'} | x \rangle e^{- \beta \bar{\cal H}(x,x')} \langle x' |
= \sum_\alpha | \alpha \rangle \lambda_\alpha \langle \alpha | \;,
\end{equation}
where the second equality represents a standard eigenvalue expansion.
By the Perron-Frobenius theorem \cite{DOMB60}, the dominant eigenvalue
$\lambda_0$ is positive and nondegenerate, and the corresponding eigenvector
$| 0 \rangle$ is the only one with elements that can all be chosen positive.
{}From ${\bf T}_0$ one can calculate
by standard methods the probability densities, correlation functions, and
partition function that fully describe the equilibrium phase \cite{DOMB60}.

The CTM method generalizes this well-known
equilibrium technique by associating with each eigenvalue
$\lambda_\alpha$ a ``constrained'' TM
${\bf T}_\alpha$ so that ``constrained'' joint and marginal probability
densities are defined in analogy with the equilibrium ($\alpha$=0) case:
\begin{eqnarray}
\label{meth-eq3}
P_\alpha(x_l,x_{l+k}) &=&
\langle\alpha|x_l\rangle\langle x_l|(\lambda_\alpha^{-1}
{\bf T}_\alpha)^{|k|}|x_{l+k}\rangle\langle x_{l+k}|\alpha\rangle \nonumber\\
P_\alpha(x_l) &=&
                  \langle\alpha|x_l\rangle \langle x_{l}|\alpha\rangle \;.
\end{eqnarray}
It was pointed out by McCraw, Schulman,
and Privman \cite{MCCR78,PRIV82A,PRIV82B} that the constrained
marginal probability densities $P_\alpha(x)$, as defined above,
can be interpreted as actual probability densities over single-layer
configurations in a constrained phase.
To obtain explicitly the constrained joint probability densities
$P_\alpha(x_l,x_{l+k})$ that define the inter-layer correlations in the
constrained phase, one must determine the constrained TM, ${\bf T}_\alpha$.
It is chosen to commute with ${\bf T}_0$, and in order to ensure convergence
of $P_\alpha(x_l,x_{l+k})$ towards stochastic independence
as $|k|$$\rightarrow$$\infty$,
each eigenvalue $| \lambda |$$>$$| \lambda_\alpha |$ is reweighted to
become $\lambda_\alpha^2/\lambda$, so that the dominant eigenvalue of
${\bf T}_\alpha$ is $\lambda_\alpha$.
A ``constrained'' free-energy density is defined by
\begin{equation}
  \label{meth-eq9}
  f_\alpha = -\frac{\ln|\lambda_\alpha|}{\beta N}+\frac{1}{\beta N}
  \sum_{x_i,x_{i+1}}P_\alpha(x_i,x_{i+1}){\rm Ln}\left(
  \frac{\langle x_i|{\bf T}_\alpha|x_{i+1}\rangle}
  {\langle x_i|{\bf T}_0 |x_{i+1}\rangle}\right)\;,
\end{equation}
where Ln($z$) is the principal branch of the complex logarithm.
For $\alpha$=0 this reduces to the standard equilibrium result,
$f = f_0 = -(N \beta)^{-1} \ln \lambda_0$.
For $\alpha$$>$0 the eigenvector $| \alpha \rangle$, corresponding to
the dominant eigenvalue of ${\bf T}_\alpha$, is orthogonal to
$| 0 \rangle$ and therefore cannot have all positive elements. As a result,
${\bf T}_\alpha$ cannot in general be a positive matrix, and
the second term in Eq.~(\ref{meth-eq9}) becomes complex-valued.
It has been observed by G{\"u}nther {\it et al.} \cite{CCAG94A}
that this second term can be considered as a complex generalization
of the Kullback discrimination function (see, {\it e.g.}, Ref.~\cite{KAPU92})
for $P_\alpha(x_l,x_{l+1})$ with respect to the divergent `probability
density' obtained by substituting ${\bf T}_0$ for ${\bf T}_\alpha$ in
Eq.~(\ref{meth-eq3}).

Successful applications of this method to calculate
$| {\rm Im} \widetilde{f} |$ have recently been performed,
both for WLRF systems \cite{GORM94,FIIG94,RIKV89,RIKV92}, and
for SRF systems \cite{CCAG93,CCAG94A,CCAG94B}.
Some results are shown in Secs.~\ref{Sec-LRF} and~~\ref{Sec-SRF}.

\section{Fluid Systems}
\label{Sec-LV}

As mentioned in Sec.~\ref{sec1}, metastable decay in fluid
systems, such as condensation of a supersaturated vapor
or freezing of a supercooled liquid, presents complications not encountered in
Ising or lattice-gas systems. In this section we mention some of these
difficulties and give a few references.

Most of the early theoretical
work on metastability that led to the formulation of CNT
[4--14]
was explicitly concerned with fluids,
as were the early mean-field approaches of van der Waals \cite{VDW1873}
and Maxwell \cite{MAXW1874}. Extensive reviews can be found in
Refs.~[17--19].

One of the most obvious differences between fluids and lattice-gas
systems is that fluids have
continuous degrees of freedom associated with droplet
translation and rotation. Their effects were considered by Lothe, Pound, and
collaborators \cite{LOTH62,FEDE66}, who evaluated the prefactor in
Eq.~(\ref{eqCNT}) and obtained an increase in the predicted condensation
rate, relative to earlier theories, on the order of 10$^{17}$.
(See, however, a recent discussion in Ref.~\cite{TALA94}.)
The long controversy over the Lothe-Pound result emphasizes
the importance and difficulty
of understanding the pre-exponential factors in the nucleation rate.

An important conceptual problem arises when one tries to formulate a precise
droplet definition
[25--29].
It was already considered by Gibbs, who introduced
a ``dividing surface'' between the liquid droplet and the vapor, but it
does not seem yet to have reached a fully satisfactory solution for fluid
systems \cite{GOUL93}.
Even in the much simpler Ising model, the proper definition of droplets has
been realized only relatively recently in terms of the
``Fortuin-Kastelyn-Coniglio-Klein-Swendsen-Wang'' cluster definition.
For general and historical discussions of the Ising droplet definition,
see Refs.~\cite{WANG89,STAU92B}, and
for discussions in the context of metastability, see
Refs.~[104--106, 145].

A serious problem of a numerical nature is the difficulty of accurately
locating the coexistence curve in a model in which its position is not
given by symmetry, as it is for the Ising or binary lattice-gas model.
(Even in more complicated lattice-gas models this
is a nontrivial problem \cite{BORG92}.)
The extremely strong dependence of the nucleation rate on the
distance from the coexistence curve will cause even a small
error in its location to produce large errors in numerical estimates of the
droplet free energy and the pre-exponential factor.

Recent large-scale computer simulations
are discussed in Refs.~\cite{SWOP90,DUIJ92},
and further reviews of progress in the theory of metastability in fluids
can be found in
Refs.~[22--24, 31].

\section{Ising Models with Weak Long-Range Forces}
\label{Sec-LRF}

In the limit of infinite interaction range, ferromagnetic Ising models with
weak, long-range interactions (WLRF models) have been rigorously shown to
possess infinitely long-lived metastable phases \cite{PENR71,PENR79,HEMM76}.
These phases lie on a hysteresis (or van~der Waals) loop that reaches
beyond the Maxwell construction in the phase diagram, and whose
critical points correspond to sharp spinodals
beyond which the metastable phases disappear.
In this respect a WLRF model is similar to a
mean-field theory \cite{VDW1873,MAXW1874,NEEL49}, except that in a
mean-field theory the only allowed form of fluctuation is a spatially
uniform change of the order parameter.  The free-energy cost of such a
change is extensive in the system volume \cite{MCCR80,PAUL89}, and the
mean-field approximation therefore predicts that metastable lifetimes
are infinite in the thermodynamic limit.

Since the behavior of a number of physical systems, including
superconductors and long-chain polymer mixtures, are often well
described by WLRF models, there has been an interest
in studying metastability in these models.  Analytic approaches
include the early field-theory treatment of a model fluid with conserved
order parameter by Cahn and Hilliard \cite{CAHI58,CAHI59},
the Fokker-Planck equation \cite{MCCR80,PAUL89,GRIF66},
renormalization-group analyses \cite{GUNT78,RIKV93}, and arguments from
random long-range bond percolation \cite{RAY90C,MONE92}, in addition to
analytic continuation of the free-energy density
[83, 92--95].
Numerical approaches include MC simulations
[129, 131, 134, 135, 153--155],
traditional TM calculations \cite{NOVO86}, and calculations of
$|{\rm Im}\widetilde f|$ by the CTM method
described in Sec.~\ref{Sec-CTM} \cite{GORM94,FIIG94,RIKV89,RIKV92}.

As pointed out by Klein and coworkers
[92, 93, 153--155],
the critical droplet in a WLRF model near the spinodal is a {\it ramified\/}
structure whose local order parameter is only slightly different from that of
the metastable phase.
(Note the contrast with the SRF case, in which the critical droplet is
compact. See Sec.~\ref{Sec-SRFn}.)
Only during the supercritical growth phase does the droplet center
compactify, leading to a measurable change in the global order parameter.
As a consequence, it is difficult to determine the
nucleation time accurately from MC simulations by monitoring the order
parameter or, equivalently, the nonequilibrium relaxation function.
Therefore, several
new techniques have been developed to analyze MC data for metastable decay in
WLRF systems, including the analysis of recrossing events \cite{PAUL88},
monitoring the number of spins in the largest cluster \cite{MONE88},
and intervention techniques \cite{MONE92}.
Some of the recent work in this area has been reviewed in Ref.\ \cite{GOUL93}.

The difficulties associated with accurately determining the nucleation rate
in WLRF models make them particularly attractive candidates for study by
alternative numerical techniques, such as the CTM method.
A simple WLRF system suitable for CTM investigation
is the Quasi-One-Dimensional Ising
(Q1DI) model, introduced by Novotny {\it et al.}\ \cite{NOVO86}.  In
its simplest form it is a chain of $L$ layers, each of which contains
$N$ Ising spins, $s_{l,n}$=$\pm$1.  Each spin interacts
ferromagnetically with each of the 2$N$ spins in the adjacent layers
with coupling $J/N$$>$0 (these are the nonzero $J_{i,j}$ in
Eq.~(\ref{eqISING}) for this model), and with an external field $H$.
In terms of
the single-layer magnetizations, $m_l$=$N^{-1}\sum_{n=1}^N s_{l,n}$,
the Hamiltonian is
\begin{equation}
\label{eqQ1DIa}
{\cal H} = -N\sum_{l=1}^L[Jm_lm_{l+1}+Hm_l]\;.
\end{equation}

\subsection{Nucleation}
\label{Sec-LRFn}

In a WLRF system like the Q1DI model, where droplet ``interfaces'' are
difficult to define, the onset of nucleation is detected by a rapid
increase in the magnetization of some layer to within a small
neighborhood of the equilibrium value.  The physical picture is that
of a bell-shaped magnetization profile that quickly develops into a
radially expanding front separating the two competing phases.  It was
shown by Gorman {\it et al.}\ \cite{GORM94} that the free-energy cost
of nucleation for the Q1DI model can be approximated by mapping the
free-energy density functional of the model to a Ginzburg-Landau form
and solving the resulting Euler-Lagrange equation.  A $\phi^3$
expansion of the potential around the exactly known \cite{NEWM80,PAUL89}
mean-field spinodal field $H_{\rm s}$, in which
$\lambda=|H|-|H_{\rm s}|\rightarrow0^-$, gives the following free-energy
cost for a $d$-dimensional WLRF system
with force range $R$ \cite{KLEI83,UNGE84}:
\begin{equation}
\label{eqQ1DIb}
  F_{\rm c} = A(T)R^d|\lambda|^{(6-d)/4}
  \left[1+O(\lambda^{1/2})\right]\;,
\end{equation}
where $A(T)$ is a nonuniversal function of $T$.  For the Q1DI
model $R$=$N$ and $d$=1 \cite{RIKV93,NOVO86}.  By solving the Euler-Lagrange
equation numerically, one can also take into account the corrections
to the expansion, as well as discrete lattice effects \cite{GORM94}.

The prefactor to the Van't Hoff-Arrhenius term in $|{\rm Im}\widetilde f|$
can be calculated by expanding the free-energy density functional
about the stationary points corresponding to the metastable phase and
the critical droplet.  The determinants of the two resulting
Schr\"odinger operators give \cite{GORM94}
\begin{equation}
\label{eqQ1DIc}
  |{\rm Im}\widetilde f| =
  B(T)(V'/V)R^{d/2}|\lambda|^{d(1-d/8)}
  \left[1+O(\lambda^{1/2})\right]
  e^{-\beta F_{\rm c}}\;,
\end{equation}
where $B(T)$ is a nonuniversal function of $T$, $V$ is the system volume, and
$V'$ is the volume of the subspace in which the droplet itself is free
to move without a cost in free energy.  For the Q1DI model,
$V'/V$=$L/(NL)$=$N^{-1}$ \cite{GORM94}.

{}From Eqs.\ (\ref{eqQ1DIb}) and (\ref{eqQ1DIc}) we see that for large
$N$, unless $H$ is extremely close to $H_{\rm s}$, the free-energy
cost $F_{\rm c}$ of surmounting the nucleation barrier is large, so
the exponential factor sets the scale for the metastable lifetime.
However, for small $N$, or for $H$$\approx$$H_{\rm s}$, the lifetime
is more strongly dependent on the particulars of the
dynamic and on the detailed
structure of the saddle point in the free-energy functional.  The
finite-range-scaling of $|{\rm Im}\widetilde f|$
near the spinodal is found by recasting
Eqs.\ (\ref{eqQ1DIb}) and (\ref{eqQ1DIc}) in terms of a scaling
variable $\zeta$=$R^{4d/(6-d)}|\lambda|$, giving
\begin{equation}
\label{eqQ1DId}
  |{\rm Im}\widetilde f| =
  (V'/V)R^{-(d/2)\left[d+3({d-2})/({6-d})\right]}\Phi(T,\zeta)
\end{equation}
with a scaling function $\Phi$ dependent only on $T$ and $\zeta$.  The
dynamic prefactor $\kappa$, taken from the lowest eigenvalue of the
Schr\"odinger operator corresponding to the saddle point, scales as
$|\lambda|^{1/2}$, but is independent of $R$.  The finite-range
scaling of the nucleation rate $\Gamma$ is thus
\begin{equation}
\label{eqQ1DIe}
  \Gamma = (V'/V)R^{-(d/2)\left[d+({3d-2})/({6-d})\right]}
  {\cal G}(T,\zeta)
\end{equation}
with a scaling function $\cal G$ dependent only on $T$ and $\zeta$.

\subsection{Numerical Transfer-Matrix Results}
\label{Sec-LRFtm}

The CTM method outlined in Sec.\ \ref{Sec-CTM} was applied by Gorman
{\it et al.}\ \cite{GORM94} to Q1DI systems of infinite length and
finite cross section $N$, with $N$ up to between 100 and
500.  Figure \ref{figQ1DIa} shows typical spectra of Re$f_\alpha$ and
$|{\rm Im}f_\alpha|$ plotted against $H$ at a fixed temperature.
These spectra are compared with the analytically continued free energy
in the limit $N$$\rightarrow$$\infty$, which corresponds to a Curie-Weiss
mean-field result (thick, dashed curves).  In each calculation, as
demonstrated in the figure, a unique $\alpha$ was found for which
Re$f_\alpha$ computed from Eq.\ (\ref{meth-eq9}) closely approximated
the real part of the mean-field metastable free energy.  Invariably,
the same $\alpha$ produced the smallest nonzero value for
$|{\rm Im}f_\alpha|$.
As $H$ was changed away from $H_{\rm s}$ towards $H$=0, the
value of the metastable $|{\rm Im}f_\alpha|$ was exponentially suppressed.

If the metastable $f_\alpha$ represents the analytically continued free-energy
density, then one expects $\ln|{\rm Im}f_\alpha|$ to be a measure of
the height of the nucleation barrier, as seen from Eq.~(\ref{eqQ1DIc}).
Extrapolated CTM estimates of
the barrier height $F_{\rm c}$ are compared in Fig.\ \ref{figQ1DIb} with the
free-energy cost of the saddle-point solution to the Euler-Lagrange
equation, which was obtained by numerical integration.
The agreement is remarkable and extends over an extremely
wide range of fields and temperatures, for most of which the lifetimes
are too long for standard Monte Carlo techniques to be useful.  Similar
results were found in a study of a three-state WLRF model \cite{FIIG94},
even where two competing metastable states
were present, suggesting that Langer's formula has a wider
applicability than previous studies have claimed \cite{GAVE89}.
These CTM results, as well as the MC results in Ref.~\cite{FIIG94}, are
consistent with Ostwald's empirical ``Law of Stages''
({\it Gesetz der Umwandlungsstufen\/}) \cite{DUNN69,OS1896},
whereby a metastable phase may decay via intermediate
metastable states before reaching equilibrium.

\section{Ising Models with Short-Range Forces}
\label{Sec-SRF}

In contrast to the situation for WLRF models, metastable
phases in systems with short-range forces (SRF models) eventually
decay, even though their lifetimes may be many orders of magnitude larger than
other characteristic timescales of the system \cite{MCDO62}. Thus, whereas
mean-field theory provides a qualitatively acceptable description of
metastability for WLRF systems in the long-range limit,
it gives a quite misleading picture for SRF systems.
The following subsections are devoted to a discussion of metastability and
metastable decay appropriate for SRF systems.

As a prototype for the metastable dynamics of SRF systems, we
consider in detail the decay of the magnetization in an impurity-free kinetic
nearest-neighbor Ising ferromagnet in an unfavorable applied field.
The critical point of this model is in the same static universality class as
the liquid/vapor phase transition \cite{GOLD92}.
(However, the liquid/vapor transition is in a different dynamic universality
class: that of Model H \cite{HOHE77}.)
The Hamiltonian is obtained
from Eq.~(\ref{eqISING}) by setting $J_{i,j}$=$J$ if $i$
and $j$ are nearest-neighbor sites and $J_{i,j}$=0 otherwise,
explicitly yielding
\begin{equation}
\label{eq1}
{\cal H} = - J \sum_{\langle i,j \rangle} s_i s_j - H \sum_i s_i
\;,
\end{equation}
where $\sum_{\langle i,j \rangle}$ and $\sum_i$ run over all
nearest-neighbor pairs and over all sites on a $d$-dimensional hypercubic
lattice of volume $L^d$, respectively \cite{NOTE}.

In a typical MC study of metastable decay in this model
one starts from the metastable phase and follows
the relaxation of the magnetization,
which is closely related to Binder's
nonequilibrium relaxation functions \cite{BIND73A}
and is directly obtained as an average
over the droplet size distribution \cite{FISH67,BIND76,STAU92}.
The volume fraction of stable phase at time $t$ is $\phi_{\rm s}(t)
= \left( m_{\rm ms} \! - \! m(t) \right) /
\left( m_{\rm ms} \! - \! m_{\rm s} \right)$,
where $m_{\rm s}$ and $m_{\rm ms}$ are the bulk equilibrium and metastable
magnetizations, respectively.
The metastable lifetime is typically estimated as the mean-first-passage-time
for $\phi_{\rm s}(t)$ to a preset value.
Monte Carlo studies using this or similar methodology have
been performed in two
[37, 104, 108, 110, 111, 120--129],
three
[38, 105, 130--132],
and higher \cite{RAY91} dimensions.
Several of these studies were analysed in terms of droplet
theory, establishing general agreement between theory and simulations.
A potentially
more accurate, but also more computationally intensive, method to
estimate the lifetimes could use the recrossing-event distribution
discussed by Paul and Heermann \cite{PAUL88}
to determine a field-dependent cutoff value for $\phi_{\rm s}$,
as suggested by Rikvold {\it et al.} \cite{RIKV94A}.

Metastability in the two-dimensional nearest-neighbor Ising ferromagnet has
also been studied by traditional TM methods by McCraw,
Schulman, and Privman \cite{MCCR78,PRIV82A,PRIV82B}, and by the CTM method by
G{\"u}nther, Rikvold, and Novotny \cite{CCAG93,CCAG94A,CCAG94B}.

\subsection{Nucleation}
\label{Sec-SRFn}

To obtain quantitative comparisons between numerical results and the
droplet-based nucleation theory, one must
calculate explicitly the field-theoretical expression
for the nucleation rate, Eq.~(\ref{eqLANG}).
This requires as input the free energy of a {\it compact\/} critical droplet
$F_{\rm c}(T,H)$, as obtained by CNT \cite{LANG67,GNW80}.
(Note the contrast with the WLRF case, in which the critical droplet is
ramified. See Sec.~\ref{Sec-LRF}.)
Sufficiently far below the critical temperature we can
calculate this by a standard droplet-theory argument
(see, {\it e.g.}, \cite{ABRA74,GUNT83A,GUNT83B}) that is modified to consider
the nonspherical droplets that appear at low $T$
because of the anisotropy of the surface tension
[108, 138, 139, 144, 160--163].
The free energy of a $d$-dimensional droplet of radius $R$
(defined as half the extent of the droplet along a primitive lattice vector)
and volume $\Omega_d R^d$ is
\begin{equation}
\label{eq5}
F(R) = \Omega_d^{(d-1)/d}  R^{d-1} \widehat{\Sigma}
     - |H| \Delta m \Omega_d R^d \;.
\end{equation}
The quantity ${\widehat{\Sigma}}$ is a temperature- and, in principle,
field-dependent proportionality factor which relates the surface
contribution to $F(R)$ with the droplet volume \cite{ZIA82,ZIA82X}.
The difference in bulk free-energy density
between the metastable and stable states is $|H| \Delta m$,
which takes some account
of droplet nesting through the magnetization difference
$\Delta m$ \cite{HARR84,BRUC81}.
Maximizing $F(R)$ yields the critical radius,
\begin{equation}
\label{eq2a}
R_{\rm c}(T,H) = \frac{(d \! - \! 1)\sigma_0}{|H| \Delta m}
\end{equation}
where $\sigma_0 \! = \! \widehat{\Sigma} / ( d \Omega_d^{1/d} ) $ is the
surface tension along a primitive lattice vector \cite{ZIA82},
and the free-energy cost of a critical droplet,
\begin{equation}
\label{eq3}
F_{\rm c}(T,H)
= \left( \frac{d \! - \! 1}{|H| \Delta m} \right)^{d-1}
  \left( \frac{\widehat{\Sigma}}{d} \right)^d \;.
\end{equation}
In addition to $R_{\rm c}$, the critical droplet is characterized by other
degrees of freedom, including the critical growth mode, droplet translations,
and deformations represented by capillary waves on the droplet surface.
The field-theoretical expression for the nucleation rate, Eq.~(\ref{eqLANG}),
properly accounts for the effects of these additional degrees of freedom,
and it is obtained explicitly by a saddle-point calculation that yields
[76--78, 90]
\begin{equation}
\label{eq4a}
\Gamma(T,H)
= A(T) |H|^{b+c} e^{- \beta F_{\rm c}(T,H) \left( 1 + O(H^2) \right)}
= A(T) |H|^{b+c}
       e^{- \left( \beta \Xi / |H|^{d-1} \right) \left( 1 + O(H^2) \right)}
\end{equation}
with
\begin{equation}
\label{eq4b}
\Xi =
\left( \frac{d \! - \! 1}{\Delta m}\right)^{d-1}
\left( \frac{\widehat{\Sigma}}{d}\right)^d
\;.
\end{equation}
The quantity $A(T)$ is a nonuniversal function of the temperature only,
\begin{equation}
\label{eq14}
b = \left\{ \begin{array}{ll} (3 \! - \! d)d/2 &
                                           \mbox{~for $1 \! < \! d \! < \! 5$,
                                                    $d \! \neq \! 3$} \\
                              -7/3     & \mbox{~for $d$=3}
            \end{array}\right.
\end{equation}
is a universal exponent related to excitations on the surface of
the critical droplet \cite{LANG67,GNW80},
and $c$ gives the $H$ dependence of the kinetic
prefactor $\kappa$ \cite{LANG68,LANG69}. The kinetic prefactor
is the only part of $\Gamma(T,H)$ that may depend explicitly on the
specific dynamic.

The Becker-D{\"o}ring cluster dynamics is defined in terms of a
master equation for the probability distribution $c_l$ of $l$-particle
clusters. The original theory only allows $l$ to change by
$\pm$1 \cite{FREN46}, but later modifications also allow
cluster coagulation and fragmentation \cite{BIND74}.
The CNT result for the exponential part of $\Gamma(T,H)$ can be obtained
from this approach. The existence and uniqueness of a solution
$c_l(t)$ that is metastable in the sense of the Penrose-Lebowitz-Sewell
criteria have been proven both for the original dynamic \cite{PENR89,KREE93A}
and for the coagulation-fragmentation generalization \cite{KREE93B}.
However, describing the clusters only in terms of their particle numbers
does not suffice to obtain the pre-exponential factor in Eq.~(\ref{eqCNT})
for a particular system. This was demonstrated explicitly by
Binder and Stauffer \cite{BIND76C}, who obtained
a result formally analogous to Eq.~(\ref{eq4a}) by using a droplet model in
which the individual clusters were characterized by several coordinates in
addition to the particle number $l$.

For $d$=2, there is substantial numerical evidence
that $b$=1, as predicted by Eq.~(\ref{eq14}). This is obtained from
calculations that do not involve the dynamics, such as analyses of
series expansions \cite{HARR84,LOWE80,WALL82} and
transfer-matrix calculations \cite{CCAG93,CCAG94A,CCAG94B}.
These studies, as well as MC work
[108, 171--173],
also indicate that the free-energy cost of the critical droplet
is given by Eq.~(\ref{eq3}) with the zero-field equilibrium values for
$\widehat{\Sigma}$ and $\Delta m$. We therefore adopt the
notations $\widehat{\Sigma}$=$\widehat{\Sigma}(T)$,
$\Delta m$=$2m_{\rm s}(T)$, and $\Xi$=$\Xi(T)$ to emphasize the lack of
field dependence in these quantities. The quantity $\widehat{\Sigma}(T)$
can be obtained
with arbitrary numerical precision by combining a Wulff construction
with the exact, anisotropic zero-field surface tension \cite{ZIA82}, and
$m_{\rm s}(T)$ is obtained
from the exact Onsager-Yang equation \cite{YANG52}.
These general results, that the surface free energy and bulk magnetization
of compact critical
droplets are determined by the zero-field equilibrium surface tension and
magnetization, respectively, are also
supported by MC studies of nucleation rates in
three dimensions \cite{RAY90B,STAU82,STAU92}.

For dynamics that can be described by a Fokker-Planck equation, it is expected
that the kinetic prefactor is proportional
to $R_{\rm c}^{-2}$ \cite{LANG68,LANG69,GNW80}, which by Eq.~(\ref{eq2a})
yields $c$=2.
This value has been confirmed numerically for the Metropolis algorithm
with updates at randomly chosen sites, but it does not appear to apply if the
sites are chosen sequentially \cite{RIKV94A}.

The numerical results cited above indicate that CNT, modified by the
post-classical results for the prefactor exponents, gives very good agreement
with the observed behavior of kinetic Ising and lattice-gas models, even
moderately far away from the coexistence curve.
A rough estimate for the field strength beyond which the simple
droplet theory discussed here should
break down (or at least become suspect) is obtained by requiring that
$2R_{\rm c}(H,T)$$>$1 \cite{CCAG94A}. The resulting crossover field
is sometimes called the ``mean-field spinodal point'',
or MFSP \cite{RIKV94A,TOMI92A}, but it is {\it not\/} identical
to the sharp spinodal found when the Ising ferromagnet is treated
in the mean-field approximation.
The MFSP is located at $H_{\rm MFSP}(T) = 2(d-1) \sigma_0(T) / 2m_{\rm s}(T)$.
The field region beyond
this limit we call {\it the strong-field region}. It will not be discussed
further here, but we hope to return to it in the future.

\subsection{Growth and the KJMA Theory}
\label{Sec-KJMA}

The reasoning behind the Kolmogorov-Johnson-Mehl-Avrami
(KJMA) theory is quite simple
[96--98, 175--178].
Critical droplets are assumed to nucleate in the metastable phase
and subsequently grow without substantial deformation \cite{NAKA60}.
(Recent discussions relevant to the limitations of the latter assumption
can be found in
Refs.~[180--183].)
The nucleation rate $\Gamma$ may either be constant (homogeneous nucleation),
or all the nuclei may already be present at $t$=0
(heterogeneous nucleation \cite{ISHI71}).
The KJMA theory is simple to work out for
both cases \cite{ISHI71,EVAN45}, but only the former will be pursued here.

The radial growth velocity, which approaches a constant limit for large
droplets, is obtained in an ``Allen-Cahn'' approximation
[72, 73, 175--178, 184--186]
as
\begin{equation}
\label{eq6}
v_\bot = (d \! - \! 1) \nu \left( R_{\rm c}^{-1} \! - \! R^{-1} \right)
\stackrel{\scriptscriptstyle R \rightarrow \infty}{\longrightarrow}
(d \! - \! 1) \nu R_{\rm c}^{-1} \equiv v_0 \;,
\end{equation}
where the coefficient $\nu$ depends on the details of the kinetics.
(To avoid confusion, we note that the usual growth law for spinodal
decomposition in systems with nonconserved order parameter, $R$$\sim$$t^{1/2}$,
results from setting $R_{\rm c}^{-1}$=0 ({\it i.e.}, $H$=0)
in Eq.~(\ref{eq6}). The $R$$\sim$$t$ growth law obtained
in the case of metastable decay is a consequence of the difference in
bulk free energy between the two phases, which acts as a driving force for
the growth \cite{GUNT83B,LIFS62,CHAN77}.)

If we set $v_\bot$=$v_0$, neglect the volumes of the critical droplets, and
consider an uncorrelated ``ideal gas'' of freely overlapping domains
of the stable phase, then the volume fraction of
stable phase at time $t$ is
\begin{equation}
\label{eqKJMA}
\phi_{\rm s}(t)
=
1 - \exp
\left[ - \Gamma {\Omega_d} v_0^d \int_0^t  (t \! - \! s)^d {\rm d}s \right]
=
1 - \exp \left[ { - \frac{\Omega_d}{d \! + \! 1}
\left( \frac{t}{t_0} \right)^{d+1} } \right] \;,
\end {equation}
where the integration variable $s$ is the time at which a particular
supercritical droplet was nucleated. This relation is known as Avrami's law.
The timescale $t_0$ is the characteristic time for collisionless
growth and sets the basic timescale for the decay of the metastable phase.
By performing the integration in Eq.~(\ref{eqKJMA}), one sees that
$t_0$ is given by
\begin{equation}
\label{eq7a}
t_0(T,H) = (v_0^d \Gamma )^{- \frac{1}{d+1}}
         = B(T) |H|^{- \frac{b+c+d}{d+1}} \exp \left[ \frac{1}{d+1}
                       \frac{ \beta \Xi (T) }{ |H|^{d-1}} \right] \;,
\end{equation}
where $B(T)$ is a nonuniversal function of $T$.
The second equality in Eq.~(\ref{eq7a})
was obtained by using Eq.~(\ref{eq4a}) for the
homogeneous nucleation rate $\Gamma(T,H)$, neglecting for simplicity the
correction term in the exponential.
By comparing Eq.~(\ref{eq7a}) for $t_0$ with Eq.~(\ref{eq4a}) for $\Gamma$,
we notice that, apart from the pre-exponential factors,
the characteristic time $t_0$ is simply given by the nucleation rate
to the power $-$$1/(d$+1) \cite{STAU92}.

Associated with $t_0$ is the characteristic lengthscale for
collisionless growth,
which we loosely call the mean droplet separation. It is given by
\begin{equation}
\label{eq7b}
R_0(T,H) = v_0 t_0
         = C(T) |H|^{- \frac{b+c-1}{d+1}} \exp \left[ \frac{1}{d+1}
                       \frac{ \beta \Xi (T) }{ |H|^{d-1}} \right] \;,
\end{equation}
where $C(T)$ is a nonuniversal function of $T$.
The lengthscale $R_0$ can be very large for weak fields, and we note that,
even though the critical droplet radius
$R_{\rm c} \! \sim \! |H|^{-1} \! \rightarrow \! \infty$ as
$|H| \! \rightarrow \! 0$,
$R_{\rm c}/R_0 \! \rightarrow \! 0$ in the same limit.

The KJMA theory gives a good approximation for sufficiently small
$\phi_{\rm s}$ (well below the percolation limit \cite{STAU92B})
that droplet correlations do not significantly alter the growth, and it
quickly became popular for analyzing
experimental results on metastable decay in a number of areas.
These include situations in which $d$ is smaller than the dimension of
the physical space and must be interpreted as the dimension of the subspace
in which the droplets grow \cite{AVRAMI}.
However, the theory does not seem to have attracted sustained
attention from theorists until the 1980's,
when Sekimoto derived exact expressions for the space-time correlation
function and the structure factor for the KJMA process
[175--178].
These results were later
generalized to infinitely \cite{AXE86} and multiply \cite{OHTA87}
degenerate equilibrium phases, and mean-field results for
the multiply degenerate case with multiple nucleation and growth rates
have been obtained, both with homogeneous and
heterogeneous nucleation \cite{BRAD89,ANDR92}.
Applications of Eq.~(\ref{eqKJMA}) have recently been made to theoretical
studies of metastable decay in long-range interaction models
and ferroelectrics
[35--38],
to nanometer-sized ferromagnetic particles \cite{RICH94},
and to kinetic Ising models \cite{DUIK90,BEAL94,RICH94,RIKV94A}.
Several of these papers contain kinetic
MC simulations \cite{DUIK90,BEAL94,RICH94,RIKV94A,AXE86,BRAD89,ANDR92}.

\subsection{Finite-Size Effects}
\label{Sec-FSE}

In an infinitely large system, the number of critical and supercritical
droplets is of course infinite, and the decay of the order parameter is
well described by Avrami's law, Eq.~(\ref{eqKJMA}).
In this section we consider how finite system size modifies the KJMA picture.
Although finite-size scaling has been extremely useful in the study of
equilibrium critical phenomena (see, {\it e.g.}, Ref.~\cite{PRIV90B}),
systematic finite-size analysis of metastable decay seems to have been
performed only recently \cite{ORIH92,DUIK90,RIKV94A,TOMI92A}.
The discussion below follows closely that of Ref.~\cite{RIKV94A}.
For simplicity we retain our exclusive focus on
hypercubic systems of volume $L^d$ with periodic boundary conditions.

For temperatures well below the critical temperature
$T_{\rm c}$ and fields inside the MFSP, the
correlation lengths in both the stable and the metastable phase are
microscopic. We are then left to consider the interplay between three lengths:
the system size $L$, the mean droplet separation $R_0$, and
the critical radius $R_{\rm c}$.

In the large-$L$ limit, where
\begin{equation}
\label{eq10}
L \gg R_0 \gg R_{\rm c} \;,
\end{equation}
the system can be approximately partitioned into
$(L/R_0)^d \! \gg \! 1$ cells of volume $R_0^d$. Each cell decays
in an independent Poisson process of rate $R_0^d \Gamma \! = \! t_0^{-1}$.
The volume fraction is then self-averaging,
so that Eq.~(\ref{eqKJMA}) can be inverted to yield the average time it takes
for the volume fraction of the equilibrium phase to increase to a
given value $\phi_{\rm s}$,
\begin{equation}
\label{eq9}
\langle t(\phi_{\rm s}) \rangle
\approx t_0(T,H)
  \left[ - \frac{d \! + \! 1}{\Omega_d}
            \ln (1 \! - \! \phi_{\rm s}) \right]^{\frac{1}{d+1}} \;.
\end{equation}
The relative standard deviation of $t(\phi_{\rm s})$ is
\begin{equation}
\label{eq10b}
r = \frac{\sqrt{\langle t(\phi_{\rm s})^2 \rangle -
          \langle t(\phi_{\rm s}) \rangle^2 }}
         {\langle t(\phi_{\rm s}) \rangle}
  \approx (R_0/L)^{\frac{d}{2}} \rho    \;,
\end{equation}
where $\rho$$\approx$1 is the relative standard deviation of a single
Poisson process (not to be confused with the lattice-gas density defined
in Eq.~(\ref{LGm})).

The regime characterized by $r \! \ll \! 1$ has been
termed ``the deterministic region'' \cite{RIKV94A,TOMI92A}.
It is subdivided
into the strong-field region mentioned in Sec.~\ref{Sec-SRFn}, and a region
characterized by a finite density of growing droplets, which we call
{\it the multi-droplet region} \cite{RIKV94A}.
Observations of the deterministic region in MC simulations are also indicated
in Refs.~\cite{DUIK90,BEAL94,RICH94,RAY90B,RIKV94A,BIND74,STOL77,STAU92}.
The characteristic absence of $L$-dependence in the multi-droplet regime
was noted by Binder and M{\"u}ller-Krumbhaar, who also derived
equations equivalent to Eqs.~(\ref{eqKJMA}) and~(\ref{eq9}) \cite{BIND74}.
Although the main emphasis was on the nucleation process, the multi-droplet
picture was also implied in Langer's work
[76--78].

For smaller $L$, so that
\begin{equation}
\label{eq11}
R_0 \gg L \gg R_{\rm c} \;,
\end{equation}
the random nucleation of a single critical droplet in a Poisson process
of rate $L^d \Gamma$ is the rate-determining step. This is
followed by relatively
rapid growth, until this droplet occupies the entire system after
an additional time much shorter than the average waiting time before a
second droplet nucleates. Therefore, the characteristic lifetime becomes
\begin{equation}
\label{eq12}
\langle t(\phi_{\rm s}) \rangle
\approx \left( L^d \Gamma(T,H) \right)^{-1}
\approx L^{-d} [A(T)]^{-1}
 |H|^{-(b+c)} \exp \left[ \frac{ \beta \Xi (T) }{ |H|^{d-1}} \right] \;.
\end{equation}
In this case $r \! \approx \! 1$, and $\langle t(\phi_{\rm s}) \rangle$
depends only weakly on the threshold $\phi_{\rm s}$.
This single-droplet region is part of ``the stochastic region''
identified in Ref.~\cite{TOMI92A}, and it was detected in MC simulations
in Refs.~\cite{DUIK90,BEAL94,RICH94,RAY90B,RIKV94A,STOL77,MCCR78,STAU92}.

The crossover between the deterministic and stochastic regimes is
determined by the condition $L \! \propto \! R_0$ with a proportionality
constant of order unity. We identify the
crossover field with the ``dynamic spinodal point'' (DSP) introduced in
Ref.~\cite{TOMI92A}, and in the limit $H \! \rightarrow \! 0$ we
explicitly obtain from Eq.~(\ref{eq7b})
\begin{equation}
\label{eq15}
H_{\rm DSP} \sim \left( \ln L \right)^{- \frac{1}{d-1}} \;.
\end{equation}
This crossover field was observed in MC simulations
reported in Refs.~\cite{DUIK90,BEAL94,RICH94,RAY90B,RIKV94A,STOL77,STAU92}.
Following Refs.~\cite{RIKV94A,TOMI92A}, we estimate $H_{\rm DSP}$ as the
field where the relative standard deviation for the lifetime is $r$=1/2.
We emphasize that, although $H_{\rm DSP}$ vanishes as
$L \! \rightarrow \! \infty$, the approach to zero
is exceedingly slow, especially for $d$=3 and above. Therefore,
$H_{\rm DSP}$ may well be measurably different from zero for
systems that are definitely macroscopic as far as their equilibrium
properties are concerned.
(As an illustration, increasing $L$ from 100 to 10$^{10}$ for $d$=3 decreases
the leading term in
$H_{\rm DSP}$ only to approximately one-half of its original value!)

Finally, we consider the small-$L$ limit,
\begin{equation}
\label{eq16}
R_0 \gg R_{\rm c} \gg L \;.
\end{equation}
In this case the volume term can be neglected in Eq.~(\ref{eq5}), and the
free-energy cost of a droplet occupying a volume fraction
$\phi_{\rm s} \! = \! V(R)/L^d$ is
$F(\phi_{\rm s}) \! \propto \! L^{d-1} \phi_{\rm s}^{(d-1)/d}
\widehat{\Sigma}(T)$ with a proportionality constant between 1 and 1/$d$,
so that the first-passage time to a given $\phi_{\rm s}$
is independent of $H$ and diverges exponentially with $L^{d-1}$.
Since the dynamics in this region of extremely weak fields
or extremely small systems is similar to that
on the coexistence line, $H$=0 \cite{WIES94},
we call it {\it the coexistence region}.
The crossover field between the coexistence and single-droplet regions,
called ``the thermodynamic spinodal point''
(THSP) in Ref.~\cite{TOMI92A}, is determined for a given $\phi_{\rm s}$ by
$\Omega_d (R_{\rm c}/L)^d \! = \! \phi_{\rm s}$, which yields
\begin{equation}
\label{eq17}
H_{\rm THSP} = \frac{1}{L \phi_{\rm s}^{1/d}}
\frac{(d \! - \! 1) \widehat{\Sigma}(T)}{2 d m_{\rm s}(T)} \;.
\end{equation}
This crossover field was observed in MC
simulations reported in Refs.~\cite{RICH94,RIKV94A,MCCR78}.

In summary, by comparing the characteristic lengths $R_0$ and $R_{\rm c}$
with the lattice constant and the system size $L$, one can identify
four different field regions, in which the
decay proceeds through different excitations. In order of increasingly strong
unfavorable field $|H|$, these are the ``coexistence region,'' characterized
by subcritical fluctuations on the scale of the system volume; the
``single-droplet region,'' characterized by decay via a single critical
droplet; the ``multi-droplet region,'' characterized by decay via a finite
density of droplets; and the ``strong-field
region,'' in which the droplet picture is inappropriate.
The crossover fields between these regions,
\begin{equation}
\label{eq20}
[ H_{\rm THSP} \! \sim \! L^{-1} ]
< [ H_{\rm DSP} \! \sim \! (\ln L)^{-\frac{1}{d-1}} ]
< [ H_{\rm MFSP} \sim L^0 ] \, ,
\end{equation}
are accurately predicted by droplet theory.
The different regions and crossover fields are illustrated in
Figs.~\ref{figA}--\ref{figC}.

In Fig.~\ref{figA} are shown average
metastable lifetimes for $L$$\times$$L$ square Ising ferromagnets with
$L$=128 and~720 at $T$=0.8$T_{\rm c}$. The data points were obtained by
the Metropolis algorithm with updates
at randomly chosen sites \cite{RIKV94A} and have here
been extrapolated past $H_{\rm THSP}$ for $L$=128 (given by Eq.~(\ref{eq17})),
using Eq.~(\ref{eq12}) for the lifetime in the
single-droplet region. The four field regions can clearly be distinguished.

An alternative
view of the information contained in Fig.~\ref{figA} is found in
Fig.~\ref{figH}, which shows as a function of $L$ the field $H_{\rm sw}$
at which $\langle t(\phi_{\rm s}$=1/2)$\rangle \! = \! \tau$, plotted for two
different values of $\tau$ at $T$=0.8$T_{\rm c}$ \cite{RICH94}.
This figure corresponds to a
contour plot of data like those shown in Fig.~\protect\ref{figA}.
In accordance with usage in the experimental
literature on small magnetic particles
[41--48],
we call $H_{\rm sw}$ the ``switching field.''
For qualitative comparison we have also included in Fig.~\ref{figH}
data digitized from Fig.~5 of Ref.~\cite{CHAN93}, which
shows the effective switching field (corrected for the demagnetization field)
{\it vs.}\ the particle diameter
for single-domain ferromagnetic barium ferrite particles, measured
by magnetic-force microscopy at room temperature.
Considering the different dimensionalities of the model and the experimental
system, and that no particular
effort was made to fit the parameters in the Ising
model to the experiments, we find the similarity between the simulated and
the experimental switching fields striking. This
qualitative agreement may indicate that the decay of the metastable
magnetization state in the barium ferrite particles proceeds through similar
nucleation and growth mechanisms as in the Ising model, and it should be
relevant to the current debate over the magnetization reversal mode in
single-domain ferromagnetic particles
[42--48].

In both the multi-droplet and the single-droplet regions, the metastable
lifetime (determined by Eqs.~(\ref{eq7a}) and~(\ref{eq12}), respectively)
has the form of an exponential in $1/|H|^{d-1}$
multiplied by a power-law prefactor in $|H|$. In both regions the derivative
of $\ln \langle t(\phi_{\rm s}) \rangle$
with respect to $1/|H|^{d-1}$ can therefore be written as
\begin{equation}
\label{eq13}
\Lambda_{\rm eff}
\equiv \frac{{\rm d} \, \ln \langle t(\phi_{\rm s}) \rangle}
            {{\rm d} \, (1/|H|^{d-1})}
= \lambda |H|^{d-1} + \Lambda \;,
\end{equation}
with $\lambda \! = \! (b \! + \! c)/(d \! - \! 1)$ and
$\Lambda \! = \! \beta \Xi(T)$ in the single-droplet region, and
$\lambda \! = \! (b \! + \! c \! + \! d)/(d^2 \! - \! 1)$
and $\Lambda \! = \! \beta \Xi(T)/(d \! + \! 1)$ in the multi-droplet region.
The quantity $\Lambda_{\rm eff}$,
calculated from the MC data in Fig.~\ref{figA}, is shown in
Fig.~\ref{figB}. The dashed straight lines correspond to Eq.~(\ref{eq13})
with the theoretically expected exponent $b$+$c$=3
and the numerically exact $\Xi(0.8T_{\rm c})$ \cite{ZIA82}
(see discussion in Sec.~\ref{Sec-SRFn}).
The data points for both system sizes follow the lower of the
two lines in the weak-field part of the
multi-droplet region. The steep rise in $\Lambda_{\rm eff}$ expected near
$H_{\rm DSP}$ is seen for both systems. However,
only the smaller one penetrates into the single-droplet region in the
field range for which data could be obtained with a reasonable amount of
computer time. A detailed statistical
analysis is given in Ref.~\cite{RIKV94A}.

Due to the dramatic increase in the metastable lifetime as $T$ is lowered,
numerical data confirming the theoretical predictions at lower temperatures
must be obtained by other techniques, such as the CTM method or
one of the new MC methods discussed in Sec.~\ref{Sec-Cmet}.
In Fig.~\ref{figC} we show
the temperature dependence of $H_{\rm DSP}$ for $L$$\times$$L$ systems
with $L$=24 and~240, as obtained by
the new MC with absorbing Markov Chains (MCAMC) method
[113--115],
together with our
analytic estimate for $H_{\rm MFSP}$. The slow decay of $H_{\rm DSP}$ with $L$
predicted by Eq.~(\ref{eq15}) can be seen. Also note the dramatic widening
of the single-droplet region as $T$ is lowered for a system of fixed size,
in agreement with recent exact predictions \cite{MART91,NEVE91,SCHO92}.

Confirmation that the quantity $\Xi(T)$ is given by its zero-field
equilibrium value is given in Fig.~\ref{figD}
(after Ref.~\cite{GORM94B}), which shows estimates of $\Xi(T)$
for $T$ between 0.17$T_{\rm c}$ ($T/J$=0.4) and 0.8$T_{\rm c}$ based on the
CTM method \cite{CCAG93,CCAG94A}, MCAMC simulations \cite{NOVO94C}, and
standard Metropolis MC simulations \cite{RIKV94A}.
The CTM estimates were obtained by fitting
the logarithmic derivative of the metastable $|{\rm Im}f_\alpha|$ to
Eq.~(\ref{eq13}) with $b$=1, $c$=0, and the correction term
from Eq.~(\ref{eq4a}) included.
The numerical estimates for $\Xi(T)$ obtained by the different methods
agree very well with each other, as well as with the exact equilibrium result.

\subsection{Discrete-Lattice Effects}
\label{Sec-DL}

Recently, a series of rigorous papers
[106, 193, 194, 196--198]
have appeared that discuss effects of the lattice
discreteness on the metastable lifetimes in the limit $T$$\rightarrow$0.
The approach is to consider a birth-death process
describing the time dependence of the fluctuations
corresponding to subcritical droplets in
the two-dimensional Ising ferromagnet with single-spin-flip
Metropolis \cite{NEVE91,SCHO92,SCOP93,KOTE93} or
modified Swendsen-Wang \cite{MART91}
dynamics. For isotropic interactions \cite{MART91,NEVE91,SCHO92,SCOP93}
and $|H|/J$$<$2 it is shown that the
metastable phase almost certainly decays through a single ``proto-critical''
droplet of spins pointing parallel to the field. This droplet is shaped like
a rectangle of $l_{\rm c}$$\times$$(l_{\rm c}$$-$1) overturned spins,
with one additional overturned spin attached as a ``knob''
to one of its long sides. The length
$l_{\rm c}$=$ \left\lceil 2J/|H| \right\rceil $ is the smallest integer
larger than $2J/|H|$, where
$2J/|H| = \lim_{T \rightarrow 0} 2 R_{\rm c}(T,H)$ is the zero-temperature
limit of the diameter of the critical droplet.
In the zero-temperature limit the average metastable lifetime is found to be
a piecewise linear function in $|H|$ which diverges asymptotically as $1/|H|$
for small $|H|$. It is given by
\begin{eqnarray}
\label{eq-lat2}
\lim_{T \rightarrow 0}
k_{\rm B} T \ln \left ( L^2 \langle t(\phi_{\rm s} \! = \! 1) \rangle \right)
&=&
8Jl_{\rm c} \! - \! \left( l_{\rm c}^2 \! - \! l_{\rm c} \! + \! 1 \right)2|H|
\nonumber \\
&\sim& 8J^2/|H| + 4J -O(|H|)
\;.
\end{eqnarray}
These results are significant for two reasons.
First, they provide an explicit dynamical justification for the existence
of a critical droplet size in kinetic Ising models.
Second, they generalize the standard continuum droplet theory
outlined in Sec.~\ref{Sec-SRFn}
to consider the discreteness of the lattice.
Using the exact zero-temperature value, $\widehat{\Sigma}(T$=$0) \! = \! 8J$,
one sees that Eq.~(\ref{eq-lat2}) is consistent to leading order
in $J/|H|$ with the continuum droplet-theory result for $\Gamma$,
Eqs.~(\ref{eq4a}) and~(\ref{eq4b}).

The temperatures at which these discrete-lattice results are expected to be
valid are so low as to be inaccessible with standard MC
techniques. However, they are
readily accessible, both with the MCAMC method
[113--115]
and with the CTM method \cite{CCAG93,CCAG94A}.
Results obtained with these methods are shown
in Fig.~\ref{figE}, together with the derivative of the first line in
Eq.~(\ref{eq-lat2}) with respect to $1/|H|$ and the corresponding
quantity in continuum droplet theory, $k_{\rm B}T \Lambda _{\rm eff}$ from
Eq.~(\ref{eq13}).
The discrete-lattice results are represented by the
series of solid, parabolic arcs, and
the continuum results for the single-droplet region
are represented by the two straight lines.
The dashed line corresponds to $b$=1 and $c$=2,
which is appropriate for the MC results, whereas
the dotted line corresponds to $b$=1 and $c$=0 and is appropriate
for the CTM results which do not contain a kinetic prefactor.
For relatively strong fields, the MC data agree reasonably well with the
discrete-lattice predictions.
For weaker fields, where the critical droplets become larger,
the oscillations caused by the lattice discreteness
become less pronounced, and the MC data points appear to approach the
continuum result.
The CTM method allows a closer approach to $H$=0 than the MC, but
the deviations from the continuum result are larger than for the MC at the
same field. A detailed
comparison of the manner in which the results obtained by the two methods
approach their respective continuum
limits, represented by the two straight lines which intersect at
the common zero-field limit $\xi$,
has yet to be performed. It is likely to involve both the difference
between the geometries of the two systems studied (square for MC and infinite
strip for CTM) and the fact that the CTM quantities can be evaluated
only at particular fields determined by the lobe structure of the metastable
$|{\rm Im}f_\alpha|$ \cite{CCAG93,CCAG94A}, which is similar to the one
shown for the WLRF Q1DI model in Fig.~\ref{figQ1DIa}.

The theoretical and numerical results discussed in this subsection
raise a number of questions that should be
answerable using the new MC and TM algorithms. In particular it is
important to understand how the lifetimes cross over to the well confirmed
results of continuum nucleation theory at higher temperatures and
weaker fields, and also how the lifetimes and the crossover
to continuum theory are affected by a change from
Metropolis to Glauber dynamics. Furthermore,
some rather dramatic effects have been predicted for the case of anisotropic
interactions as $T$$\rightarrow$0 \cite{KOTE93}. No sign of these effects
were found in a recent study by the CTM method \cite{CCAG94B},
and a further investigation by the MCAMC method is in progress.

\section{Summary and Discussion}
\label{Sec-D}

In this review we
have presented some aspects of the current knowledge regarding the
mechanisms and rates of decay of metastable phases in statistical-mechanical
systems. With its record of 270 years of published research in a variety
of basic and applied contexts following Fahrenheit's 1724 paper \cite{F1724},
this is indeed a venerable field of inquiry.
Yet we believe it is a sign of continued vigor and relevance that nearly half
of the references we cite have appeared within the last decade.

It is now well understood that metastable decay is a kinetic phenomenon,
whose rate depends on the interplay between the (homogeneous and/or
heterogeneous) nucleation of critical droplets of the equilibrium phase and
on the subsequent growth of the supercritical droplets.
Formally, the nucleation rate is given by the Van't Hoff-Arrhenius relation
of classical nucleation theory, Eq.~(\ref{eqCNT}), or the
field-theoretical relation, Eq.~(\ref{eqLANG})
[76--78],
and the simultaneous nucleation and growth give rise to ``Avrami's law'',
Eq.~(\ref{eqKJMA})
[96--98].
However, both the free energy $F_{\rm c}$ of the critical droplet and
the pre-exponential factor $\cal P$ in Eq.~(\ref{eqCNT})
(both of which are contained in the imaginary part of the metastable free
energy, Im$\widetilde{f}$, in Eq.~(\ref{eqLANG})) depend crucially on the
geometry of the critical droplet. Whereas the shape of the critical droplet
determines the exponential factor in Eq.~(\ref{eqCNT}), the prefactor
is determined by the fluctuations around the critical droplet shape. The
latter is evaluated in the ``post-classical'' field-theoretical approach by
a steepest-descent calculation around a saddle point representing the
critical droplet.

If the order parameter is constrained to be uniform over the entire system,
so that $F_{\rm c}$ becomes extensive in the system size, one recovers the
mean-field picture of van~der Waals \cite{VDW1873}, Maxwell \cite{MAXW1874},
and N{\'e}el \cite{NEEL49}. In this picture the
metastable phases are infinitely long-lived in the limit of infinite system
size, and the limit of metastability is marked by a sharp spinodal line.

A different class of systems are the weak-long-range-force (WLRF) models
discussed in Sec.~\ref{Sec-LRF}, which may be useful models for,
{\it e.g.}, superconductors, long-chain polymers, and systems with elastic
interactions. These models also have infinitely long-lived
metastability and a sharp spinodal
in the limit of infinite interaction range \cite{PENR71,PENR79,HEMM76}.
However, the critical droplet is a ramified object
[92, 93, 153--155]
whose free energy is given in terms
of the interaction range and the distance from the spinodal by
Eq.~(\ref{eqQ1DIb}) \cite{KLEI83,UNGE84},
and the resulting nucleation rate is given by Eq.~(\ref{eqQ1DIe}).
The agreement between these analytical results and the numerical
constrained-transfer-matrix (CTM) method is shown in
Fig.~\ref{figQ1DIb} \cite{GORM94},
which illustrates the dramatic increase in metastable lifetime as the distance
from the mean-field spinodal increases.

For short-range-force (SRF) models, which are useful to represent, {\it e.g.},
anisotropic magnets governed by exchange interactions, and which belong to
the same static universality class as the liquid/vapor phase transition,
the situation is again different. In this case the critical droplet is a
compact object with a radius $R_{\rm c}$ given
by Eq.~(\ref{eq2a}), and $F_{\rm c}$ is given by Eq.~(\ref{eq3}).
The nucleation rate from the
field-theoretical saddle-point calculation is given in
Eqs.~(\ref{eq4a})--(\ref{eq14})
[76--78, 90],
and the effects of
simultaneous nucleation and growth are reflected in the explicit form of
Avrami's law, Eqs.~(\ref{eqKJMA}) and~(\ref{eq7a})
[96--98].
In contrast to mean-field and WLRF systems,
SRF systems do {\it not\/} have a sharp spinodal.

Finite-size effects represent an aspect of metastable kinetics which has
attracted increased attention
in recent years \cite{ORIH92,DUIK90,RIKV94A,TOMI92A}. For SRF systems, these
effects give rise to different dependences of the metastable lifetime
on applied field (or chemical
potential, pressure, or supersaturation) and system size for different
values of these parameters. These regions are separated by {\it size
dependent\/} crossover fields:
the thermodynamic spinodal field $H_{\rm THSP}$ given in Eq.~(\ref{eq17}),
the dynamic spinodal field $H_{\rm DSP}$ given in Eq.~(\ref{eq15}),
and the size-{\it in\/}dependent ``mean-field'' spinodal field $H_{\rm MFSP}$.
For $H_{\rm THSP} \! < \! |H| \! < \! H_{\rm DSP}$ the decay occurs through
a {\it single\/} droplet, and the mean lifetime is given by
the inverse nucleation rate through Eq.~(\ref{eq12}).
In contrast, for $H_{\rm DSP} \! < \! |H| \! < \! H_{\rm MFSP}$
the decay involves
a {\it nonzero density\/} of simultaneously nucleating and growing
droplets, and the mean lifetime is given by
Avrami's law through Eqs.~(\ref{eq7a}) and~(\ref{eq9}).
The detailed agreement between these analytical results and
Monte Carlo (MC) simulations is illustrated in
Fig.~\ref{figA}, which shows the dramatic dependence on 1/$|H|^{d-1}$
of the logarithm of the average
metastable lifetime for two-dimensional systems of
different sizes. Similar agreement has also been demonstrated for
three-dimensional systems
[130--132],
and a rigorous justification for the nucleation-theory results has been
obtained in two dimensions at very low temperatures
[106, 193, 194, 196--198].
Further agreement between analytic theory and simulation results is
illustrated in Figs.~\ref{figB}, \ref{figD}, and~\ref{figE}.

A number of questions regarding metastable decay still remain to be answered.
Probably some of the most difficult ones are related to fluids,
which we discussed briefly in Sec.~\ref{Sec-LV}.
In particular we would like to see developed a droplet definition as
unambiguous as the ``Fortuin-Kastelyn-Coniglio-Klein-Swendsen-Wang''
definition
[104--106, 145--147]
currently used for Ising and lattice-gas systems.

Some of the most exciting potential for both fundamental and
technological progress may lie in combining modern ``atomic engineering''
techniques, such as atomic and magnetic force microscopies, with
computer simulation. Using these experimental techniques, one could
both build and study well-characterized metastable systems. Employing
state-of-the-art computer hardware and algorithms, one could simulate
these same systems with tens to billions \cite{ITO93,STAU92} of particles.
This approach should provide the opportunity to
study effects of size, boundary conditions, and impurities by a combination
of theory, simulation, and experiment in a carefully controlled manner.
A step in this direction was recently taken by Richards
{\it et al.}\ \cite{RICH94}, some of whose results are shown together with
experimental data from Ref.~\cite{CHAN93} in Fig.~\ref{figH}.

In summary, we think the field of metastability is a healthy 270-year-old with
a long, exciting life ahead of it.

\centerline{\bf Acknowledgments}

P.A.R.\ dedicates this article to the memory of his father,
Per Rikvold (1919--1994).

We gratefully acknowledge discussions with M.~A.\ Novotny,
C.~C.~A.\ G{\"u}nther, H.~L.\ Richards, and J.~Lee,
correspondence with J.~Lothe, K.~Nishioka,
H.~Reiss, and D.~Stauffer, and permission to use published and unpublished
data by M.~A.\ Novotny, C.~C.~A.\ G{\"u}nther,
H.~L.\ Richards, and S.~W.\ Sides.
This work was supported by Florida State University through the Supercomputer
Computations Research Institute (U.S.\ Department of Energy Contract No.\
DE-FC05-85ER25000) and the Center for Materials Research and Technology,
and by the U.S.\ National Science Foundation Grants
No.~DMR-9013107 and~DMR-9315969.

\clearpage

\begin{thebibliography}{100}
\bibitem{ZIA82X}
Our quantity $\widehat{\Sigma}(T)$ is identical to $dW^{1/d}$ in Eq.~(3) of
  Ref.~\cite{ZIA82}.
\end{thebibliography}

\clearpage

\begin{figure}
\caption{
The branches of ${\rm Re}f_\alpha$ (top) and $|{\rm Im}f_\alpha|$ (bottom)
computed from constrained transfer matrices for
a 35$\times$$\infty$ Q1DI system at $T$=$0.5T_{\rm c}$ and plotted
versus $H$ in units of $\beta_{\rm c}^{-1}$$\equiv$$T_{\rm c}$=2$J$.
The equilibrium branch of $|{\rm Im}f_\alpha|$ is
identically zero and has been removed for clarity.  The thick dashed
lines represent the mean-field values for the real and imaginary parts
of the analytically continued free energy. The dotted vertical line
indicates the mean-field spinodal field $H_{\rm s}$.
After Ref.~\protect\cite{GORM94}.
}
\label{figQ1DIa}
\end{figure}

\begin{figure}
\caption{
Extrapolated finite-$N$ CTM estimates for $F_{\rm c}/N$ in
Eq.\ (\protect\ref{eqQ1DIb}), plotted in units of
$\beta_{\rm c}^{-1}$$\equiv$$T_{\rm c}$=2$J$
against $|\lambda|/H_{\rm s} = (H_{\rm s}-H)/H_{\rm s}$
for various temperatures $T$.  The lines
indicate the free-energy cost of forming the saddle-point solution of
the Euler-Lagrange equation at the same temperatures and were
obtained by numerical integration.  After Ref.~\protect\cite{GORM94}.
}
\label{figQ1DIb}
\end{figure}

\begin{figure}
\caption{
Metastable lifetimes for $L$$\times$$L$ two-dimensional
Ising ferromagnets with $L$=128
($\circ$) and~720 (solid diamonds), shown on a logarithmic scale {\it vs.}\
$J/|H|$. The data were obtained by the Metropolis
algorithm with updates at randomly chosen sites
for $T$=0.8$T_{\rm c}$. The lifetimes, which are
given in units of MC steps per spin (MCSS), were
estimated as the mean-first-passage times to
$m$=0.7 ($\phi_{\rm s}$$\approx$0.13) \protect\cite{RIKV94A}.
The solid curves represent
extrapolations to $|H|$$<$$H_{\rm THSP}$ for the smaller system,
using Eq.~(\protect\ref{eq12}) for the lifetime in the single-droplet region.
$H_{\rm THSP}^{-1}$ is given by Eq.~(\protect\ref{eq17})
and is indicated by the arrow marked THSP(128).
The other arrows mark the mean-field spinodal (MFSP) for both systems
and the dynamic spinodal points (DSP) for $L$=128 and~720 respectively.
The field-regions seen in the figure are, from left to right,
the strong-field region, the multidroplet region,
the single-droplet region, and the coexistence region (the latter for
$L$=128 only).
The dashed lines correspond to the asymptotic slopes in the single-droplet
and multidroplet regions, $\beta \Xi$ and $\beta \Xi /3$, respectively.
Data courtesy of S.~W.\ Sides.
}
\label{figA}
\end{figure}

\begin{figure}
\caption{
The switching field $H_{\rm sw}(L,\tau)/J$ (left-hand vertical axis)
shown {\it vs.}\ $L$ (bottom horizontal axis)
for waiting times $\tau$=100~MCSS ($\circ$)
and~1000~MCSS ($\times$) in an $L$$\times$$L$ Ising ferromagnet
at $T$=0.8$T_{\rm c}$. The data were obtained by
MCAMC simulations, and the error bars, everywhere much smaller than the
symbol size, are not shown.
This figure corresponds to a contour plot of lifetime data like those shown in
Fig.~\protect\ref{figA}.
The dotted curve represents $H_{\rm THSP}$ and the dashed curve represents
$H_{\rm DSP}$. To the left of $H_{\rm THSP}$
is the coexistence region (CE), between the
two spinodals is the single-droplet region (SD), and to the right of
$H_{\rm DSP}$ is the multidroplet region (MD).
Data courtesy of H.~L.\ Richards and S.~W.\ Sides \protect\cite{RICH94}.
For qualitative comparison we have also included data digitized from Fig.~5 of
Ref.~\protect\cite{CHAN93} ($\Diamond$ with error bars),
showing the effective switching field
(right-hand vertical axis) {\it vs.}\ the
particle diameter (top horizontal axis)
for single-domain ferromagnetic barium ferrite particles.
}
\label{figH}
\end{figure}

\begin{figure}
\caption{
Here are shown {\it vs.}\ $|H|/J$ estimated values of the quantity
$k_{\rm B}T \Lambda_{\rm eff}/J^2$ from Eq.~(\protect\ref{eq13}),
based on the data from Fig.~\protect\ref{figA}.
The upper and lower dashed straight lines correspond to the
theoretical result in the single-droplet
and in the multidroplet region, respectively.
Both were calculated using the
theoretically expected exponent $b$+$c$=3 and the numerically
exact $\Xi(0.8T_{\rm c})/J^2$$\approx$$0.918 \, 915$.
The pairs of opposing arrows with error-bar ``feathers'' mark $H_{\rm DSP}$
for $L$=128 (right pair) and~720 (left pair).
See detailed discussion in Sec.~\protect\ref{Sec-FSE}.
After Ref.~\protect\cite{RIKV94A}.
}
\label{figB}
\end{figure}

\begin{figure}
\caption{
The dynamic spinodal field, $H_{\rm DSP}/J$ {\it vs.} $k_{\rm B}T/J$
for $L$$\times$$L$ systems
with $L$=24 ($\circ$ with error bars) and~240 ($\times$ with error bars),
obtained by the MCAMC algorithm.
The dotted curves are merely guides to the eye.
Data courtesy of M.~A.\ Novotny \protect\cite{NOVO94C}. Also shown is our
analytic estimate for the $L$-independent $H_{\rm MFSP}/J$ (solid line).
}
\label{figC}
\end{figure}

\begin{figure}
\caption{
The quantity $\Xi(T)/J^2$ as obtained from the CTM method ($\circ$)
\protect\cite{CCAG93,CCAG94A}, MCAMC simulations (solid diamond)
\protect\cite{NOVO94C}, and
Metropolis simulations (solid square) \protect\cite{RIKV94A},
shown {\it vs.} $k_{\rm B}T/J$.
The solid curve corresponds to
the zero-field equilibrium value of $\Xi(T)$ \protect\cite{ZIA82}.
The corresponding critical-droplet shape interpolates between a square at
$T$=0 and a circle at $T$=$T_{\rm c}$, both shown for the whole temperature
range as dashed curves.
See detailed discussion in Sec.~\protect\ref{Sec-FSE}.
After Ref.~\protect\cite{GORM94B}.
}
\label{figD}
\end{figure}

\begin{figure}
\caption{
Here are shown {\it vs.}\ $|H|/J$ estimated values of the quantity
$k_{\rm B}T\Lambda_{\rm eff}/J^2$ from Eq.~(\protect\ref{eq13}),
based on metastable lifetimes for a 24$\times$24 Ising ferromagnet at
$T$=0.4$J$ as obtained by the MCAMC method \protect\cite{NOVO94A,NOVO94B}
($\circ$ with error bars), and on the metastable ln$|{\rm Im}f_\alpha|$ for
a 9$\times$$\infty$ system at the same temperature, obtained by the CTM
method \protect\cite{CCAG94A} ($\Diamond$).
The theoretically expected results from continuum droplet theory as given by
the rhs of Eq.~(\protect\ref{eq13}) are represented by straight lines.
The dashed line corresponds to $b$=1 and $c$=2,
appropriate for the MC results, and the dotted line to $b$=1 and $c$=0,
appropriate
for the CTM results, which do not contain a kinetic prefactor. Both lines
intersect the $y$-axis at the numerically exact $\Xi(T$=0.4$J)/J^2$.
An estimate of the effects of the lattice discreteness is given by the
solid parabolic arcs, which represent the $1/|H|$ derivatives of the first
line in Eq.~(\protect\ref{eq-lat2}).
Data courtesy of M.~A.\ Novotny and C.~C.~A.\ G{\"u}nther.
After Ref.~\protect\cite{NOVO94B}.
}
\label{figE}
\end{figure}

\end{document}